\documentclass[prd,preprint,a4paper,superscriptaddress,nofootinbib,12pt]{elsarticle}
\usepackage{amssymb}
\usepackage{graphicx}
\usepackage{amsmath}
\usepackage{eurosym}
\usepackage{dsfont}
\usepackage{amsmath}
\usepackage{amssymb}
\usepackage{amsthm}
\usepackage{stmaryrd}
\usepackage{mathrsfs}
\usepackage{textcomp}
\usepackage{amsfonts}
\usepackage{txfonts}
\usepackage[T1]{fontenc}
\usepackage{color}

\newcommand{\ds}{\displaystyle}
\newcommand{\eso}{\ensuremath{\mathfrak {so}}}
\newcommand{\gel}{\ensuremath{\mathfrak{gl}}}
\newcommand{\hM}{\ensuremath{\hat{M}}}
\newcommand{\mE}{\ensuremath{{\mathcal E}}}
\newcommand{\hMmn}{\ensuremath{\hat{M}_{\mu\nu}}}
\newcommand{\hx}{\ensuremath{\hat{x}}}
\newcommand{\ieso}{\ensuremath{\mathfrak {iso}}}
\newcommand{\igel}{\ensuremath{\mathfrak {igl}}}
\newcommand{\ka}{\ensuremath{\kappa}}
\newcommand{\Mmn}{\ensuremath{M_{\mu\nu}}}

\def\U{\mathcal{U}}

\begin{document}

\title{{\Large Generalized Poincar\'{e} algebras, Hopf algebras and \ka-Minkowski
  spacetime}}
\author{D. Kova\v{c}evi\'{c}  \footnote{domagoj.kovacevic@fer.hr}}
\address{Faculty of Electrical Engineering and Computing, Unska 3, HR-10000
Zagreb, Croatia}
\author{S. Meljanac \footnote{meljanac@irb.hr}}
\address
{Rudjer Bo\v{s}kovi\'{c} Institute, Bijeni\v{c}ka c.54, HR-10002 Zagreb, Croatia}
\author{A. Pacho{\l} \footnote{pachol@raunvis.hi.is}}
\address{Science Institute, University of
Iceland, Dunhaga 3, 107 Reykjavik, Iceland}
\author{R. \v{S}trajn  \footnote{rina.strajn@gmail.com}}
\address
{Rudjer Bo\v{s}kovi\'{c} Institute, Bijeni\v{c}ka c.54, HR-10002 Zagreb, Croatia}

\begin{abstract}
We propose a generalized description for the \ka-Poincar\'{e}-Hopf algebra as a symmetry quantum
group of underlying \ka-Minkowski spacetime. We investigate all the possible implementations
of (deformed) Lorentz algebras which are compatible with the given choice of \ka-Minkowski algebra realization. For the given realization of \ka-Minkowski spacetime there is a unique \ka-Poincar\'{e}-Hopf algebra with undeformed Lorentz algebra. We have constructed a three-parameter family of deformed Lorentz generators with \ka-Poincar\'{e} algebras which are related to \ka-Poincar\'{e}-Hopf algebra
with undeformed Lorentz algebra. Known bases of \ka-Poincar\'{e}-Hopf algebra are obtained as
special cases.
 Also deformation of \igel (4) Hopf algebra  compatible with the \ka-Minkowski spacetime is presented. Some physical applications are briefly discussed. 
\end{abstract}

\maketitle

\section{Introduction}

The Poincar\'{e} symmetry is the full symmetry of special relativity and it includes
translations, rotations and boosts. Algebraically it is described by the Poincar\'{e}-Lie algebra (Lorentz generators $M_{\mu\nu}$ and momenta $P_{\mu}$), usually
denoted as $\mathfrak{iso}(1,3)$, and defined by the following commutation relations
\begin{gather}  \label{isog1}
  [\Mmn,M_{\lambda\rho}]=i\left(
    M_{\mu\lambda}\eta_{\nu\rho}-M_{\mu\rho}\eta_{\nu\lambda}-
    M_{\nu\lambda}\eta_{\mu\rho}+M_{\nu\rho}\eta_{\mu\lambda}\right), \\
  [M_{\mu \nu },P_{\lambda }]= i\left(P_\nu\eta_{\mu \lambda}-
    P_\mu\eta_{\nu\lambda}\right), \\
    [P_{\mu},P_{\nu}]=0,
\end{gather}
where $\eta_{\mu\nu}$ is the metric tensor with Lorentzian signature. As it is known,
any Lie algebra provides an example of undeformed Hopf algebra via its universal
enveloping
algebra. Therefore, the universal enveloping algebra of the Poincar\'{e}-Lie algebra
$\U_{\mathfrak{iso}(1,3)}$  can be equipped with comultiplication, a counit and
antipode
\begin{equation}
\Delta_0M_{\mu\nu }=M_{\mu\nu }\otimes 1+1\otimes M_{\mu\nu }\quad
\mbox{and}\quad\Delta_0P_\mu=P_\mu\otimes 1+1\otimes P_\mu
\end{equation}
\begin{equation}\label{SPoin}
S(M_{\mu\nu })=-M_{\mu\nu };\qquad S(P_\mu)=-P_\mu;\qquad
\epsilon(M_{\mu\nu })=\epsilon(P_\mu)=0
\end{equation}
 defined on the generators and then extended to the whole $\U_{\mathfrak{iso}
(1,3)}$, which constitutes the undeformed Poincar\'{e}-Hopf algebra.
Such an algebra can be deformed within the Hopf algebraic framework. First
deformations of Poincar\'{e} symmetry appeared in the early 90's. The so-called
$\kappa$-deformation \cite{Luk1,MR}, was obtained by contraction procedure from
q-deformed $SO_q(3,2)$. It gained much attention since such deformed relativistic
symmetries can be interpreted as a description of Planck scale world and the
deformation parameter $\kappa$ (of mass dimension) is usually interpreted as the
Planck Mass $M_P$ or the quantum gravity scale $M_{QG}$. The
$\kappa$-Poincar\'{e}-Hopf algebra is the symmetry algebra (quantum group) of
underlying quantum space which should replace the undeformed one at this
level of energy. It is believed that below the Planck scale a more general spacetime
structure appears, e.g. noncommutative one, where (as with quantum
mechanics phase space) uncertainty relations naturally arise \cite{Dop1}.
The $\kappa$-Minkowski spacetime, as one of the examples of noncommutative
spacetime is a (Hopf) module algebra over the $\kappa$-Poincar\'{e}-Hopf algebra
and it stays invariant under the quantum group of transformations in analogy to the
classical case. The $\kappa$-deformed Poincar\'{e} algebra as deformed symmetry
of the $\kappa$-Minkowski spacetime inspired many authors to, e.g. construct
quantum field theories (see e.g., \cite{klm00,klry09,ms11,mstw11}), electrodynamics on
$\kappa$-Minkowski spacetime \cite{hjm11,dj11}, or modify particle statistics
(see e.g., \cite{gghmm08,kappaSt}). One of the main ideas is to discuss Planck
scale (quantum gravity) effects,  since it naturally includes $\kappa$ (Planck scale,
quantum gravity) corrections.

In this Letter we are interested in the generalization of the description for the
$\kappa$-deformed Poincar\'{e} symmetry group, especially the Lorentz sector.
A deformation of the phase space, compatible with $\kappa$-Minkowski spacetime
can be characterized by a set of functions $h_{\mu\nu}(p)$ \cite{k-jm11,km11,
glik02,k-gn02}. For a given choice of $h_{\mu\nu}$ there are infinitely many ways
of implementing the Lorentz algebra, with different Hopf algebra structures generally within 
$\mathfrak{igl}(4)$. Among these algebras, for a given choice of $h_{\mu \nu}$,
there exists a unique  $\kappa$-Poincar\'{e}-Hopf algebra. Such examples were
considered in \cite{MR,k-jm11,km11,glik02,k-gn02}. The first example of
deformed Lorentz algebra with the $\kappa$-Poincar\'{e}-Hopf algebra was proposed
in \cite{Luk1}. Undeformed Lorentz algebras that cannot be equipped with a Hopf
algebra structure which would be closed in the Poincar\'{e} algebra were considered
in \cite{klry09,gghmm08,bklvy,klry08,bp09,byk09}. Our aim is to introduce the most
general form of deformation of the Lorentz algebra, compatible with the
$\kappa$-Minkowski spacetime, which includes the above examples as special
cases. Especially, we have constructed an infinite three-parameter class of deformed
Lorentz algebras with the $\kappa$-Poincar\'{e}-Hopf algebra structure which are
related to the $\kappa$-Poincar\'{e}-Hopf algebra with undeformed Lorentz algebra, of
 which the standard basis \cite{Luk1} is one example.

The plan of the Letter is as follows. In Section \ref{dha} we introduce the deformed
Heisenberg algebra, including noncommutative $\kappa$-Minkowski spacetime
coordinates. Later, in Section \ref{gpa} we focus on the generalization of
$\kappa$-Poincar\'{e} algebra which includes a certain ansatz for Lorentz
boost generators. This section constitutes the generalized description of the
deformed Poincar\'{e} algebra within this framework. In Section \ref{lapha} we focus
on the undeformed Lorentz algebra with a deformed coalgebra which depends
on the choice of $\kappa$-Minkowski realization. Here the bicrossproduct basis
\cite{MR} for the $\kappa$-Poincar\'{e}-Hopf algebra is included as a special
case. Section \ref{dla} includes the case in which the Lorentz algebra is deformed
and here we are able to obtain the standard basis \cite{Luk1} of
$\kappa$-Poincar\'{e} as a special case. Section \ref{mst} concerns the twist
deformation of $\mathfrak{igl}(4)$. Finally, in Section \ref{conc} conclusions are
presented.

\section{Deformed Heisenberg algebra}
\label{dha}

Before going into deformed relativistic symmetries let us start with the deformation
of a phase space (Heisenberg algebra) including $\kappa$-Minkowski spacetime.
In the undeformed case Heisenberg algebra $\ H$ can be defined as a free
algebra of $n$ coordinate generators and $n$ generators of momenta,
satisfying the following relations:
\begin{equation}
\left[ x_{\mu },x_{\nu }\right] =0;\qquad \left[ p_{\mu },x_{\nu } \right]
=-i\eta _{\mu \nu }\cdot 1;\qquad \left[ p_{\mu },p_{\nu }\right] =0;
\end{equation}
where $\eta_{\mu\nu}=(-,+,+,+)$ is diagonal metric tensor with Lorentzian
signature.\\
A possible deformation of Heisenberg algebra is to introduce noncommutative
coordinates. In the case of $\kappa -$Minkowski spacetime algebra we have:
\begin{equation}
  \left[\hat{x}_\mu,\hat{x}_\nu\right] =i\left(a_\mu\hat{x}_\nu-a_\nu\hat{x}_\mu
  \right);\qquad
  \left[ p_{\mu },\hat{x}_{\nu }\right]=-ih_{\mu \nu }\left( p\right);\qquad
  \left[ p_{\mu },p_{\nu }\right] =0.
  \label{b3}
\end{equation}
For simplicity we can choose $a_\mu=(a_0,0,0,0)$.
We shall denote the algebra described by these commutation relations as $\hat{%
H}$. The choice of function $h_{\mu\nu }\left( p\right) $ is such that the
Jacobi identities for $\hat{x}_{\mu },p_{\nu }$\ are satisfied which implies
that $h_{\mu \nu }\left( p\right) $ obeys a system of partial differential
equations (\cite{k-jm11,km11}, see also \cite{glik02,k-gn02,k-jms07,ms06,bp11}).
In the undeformed limit $a_{\mu }\rightarrow 0$ we obtain algebra $H$%
, i.e. $\lim_{a_0\rightarrow 0}h_{\mu\nu}=\eta _{\mu\nu }$. 
For the purpose of deformation we shall introduce h-adic extension of $\hat{H}$
and denote it by $\hat{H}[[a]]$, which is an algebra of formal power series
\cite{bp11}.
In fact, algebra $\hat{H}[[a]]$ can be obtained from $H$ by non-linear change of
generators (mapping), as example see realization (\ref{realZ}) \cite{bp11}.

Let us emphasize that for the given $\hat{H}$, there is a unique choice of
functions $h_{\mu\nu}$ and the corresponding realization, $\hat{x}_\mu=
x^\alpha h_{\alpha\mu}(p)$. One family of realizations can be described by:
\begin{equation}\label{real}
\hat{x}_{0}=x_{0}\psi\left(A\right) -a_{0}x_{k}p_{k}\gamma \left( A\right),\;\;\;
\hat{x}_{i}=x_{i}\varphi \left( A\right)
\end{equation}
where $A=-a_\alpha p^\alpha=a_0p_0$.
The functions $\varphi \left( A\right) ,\psi \left( A\right) $ are arbitrary
real-analytic functions such that $\varphi \left( 0\right) =1;\psi \left(
0\right) =1$ and $\varphi ^{\prime }\left( 0\right) $ is finite (where $\varphi
^{\prime }=\frac{d\varphi }{dA}$) and $\gamma \left( A\right) $ obeys:
$\gamma=\frac{\varphi'}{\varphi}\psi+1$, in order to obtain
$\kappa-$Minkowski commutation relations for noncommutative coordinates
\cite{k-jm11,ms06}. Functions $h_{\mu\nu}$ can be easily obtained from
relations (\ref{b3}) and (\ref{real}).

As a special case of Eq. (\ref{real}) one can choose: $\psi=1;\;\varphi =
Z^{-\lambda }$. Then we have $\gamma=1-\lambda$, $Z=e^A$ and
\begin{equation}\label{realZ}
\hat{x}_{0}=x_{0}-a_{0}\left(1-\lambda \right) x_{k}p_{k},\;\;\;
\hat{x}_{i}=x_{i}Z^{-\lambda}
\end{equation}
where we introduced the so-called shift operator $Z\in \hat{H}[[a]]$ defined
by commutation relations: $\left[ Z,\hat{x}_{\mu }\right] =ia_{\mu
}Z;\, \left[ Z,p_{\mu }\right] =0$.

Note that the above realizations are not Hermitian, however one can construct
the Hermitian realizations by simple formula: 
\begin{equation}
\hat{x}_{\mu }^{h}=\frac{\hat{x}_{\mu }+\hat{x}_{\mu }^{\dagger }}{2}.
\end{equation}

\section{Generalized Poincar\'{e} algebras}
\label{gpa}

We are interested in the most general form for deformation of Poincar\'{e} algebra
which would be
compatible with the above deformed Heisenberg algebra $\hat{H}$.
Let us define deformed Lorentz generators \hMmn\ satisfying general
commutation relations
\begin{equation}
  [\hMmn,\hM_{\lambda\rho}]=i\left(
    \hM_{\mu\lambda}g_{\nu\rho}(p)-\hM_{\mu\rho}g_{\nu\lambda}(p)-
    \hM_{\nu\lambda}g_{\mu\rho}(p)+\hM_{\nu\rho}g_{\mu\lambda}(p)\right)
  \label{c4}
\end{equation}
where $g_{\mu\nu}(p)$ is real, nondegenerate metric in
momentum space $g_{\mu\nu}(p)=g_{\nu\mu}(p)$ and \hMmn\ satisfy
$\hMmn=-\hM_{\nu\mu}$. This point of view could be related with the recent
idea of relative locality \cite{rel_loc} which includes the description of
$\kappa$-Poincar\'{e}-inspired momentum space geometry. However we will not
focus on this here.

We write \hMmn\ in order to emphasize the deformation. The generalized Poincar\'{e} algebra is defined by $[p_\mu,p_\nu]=0$ and
\begin{equation}
  [\hM_{\mu\nu},p_\lambda]=iG_{\mu\nu\lambda}(p)
  \label{c6}
\end{equation}
for real functions $G_{\mu\nu\lambda}(p)$.
We require that all Jacobi
identities are satisfied which set restrictions on tensors $g_{\mu\nu}$,
$G_{\mu\nu\lambda}$.
It is enough to calculate Jacobi identities using Eq. (\ref{c4}) for three \hM s and
using Eqs. (\ref{c4}) and (\ref{c6}) for two \hM s and for one $p$.
In order to complete the set of commutation relations, we include
\begin{equation}
  [\hM_{\mu\nu},\hx_\lambda]=-i\hx_\alpha K_{\mu\nu\lambda}^\alpha(p)
  \label{c7}
\end{equation}
for real functions $K_{\mu\nu\lambda}^\alpha(p)$.
Since $\hat{x}_\mu=x^\alpha h_{\alpha\mu}(p)$ satisfy \ka-Minkowski
commutation relations  (\ref{b3}) and $\hMmn=x^\alpha G_{\mu\nu\alpha}(p)$
(satisfying Eqs. (\ref{c4}) and (\ref{c6})), Eq. (\ref{c7}) produces expressions
for functions $K_{\mu\nu\lambda}^\alpha$ in terms of $h_{\alpha\mu}$ and
$G_{\mu\nu\alpha}$. We also require the smooth limit to the Poincar\'{e} algebra
when deformation parameter $a_\mu$ goes to 0.

The deformed Heisenberg algebras are introduced as subalgebras of the generalized Poincar\'{e} algebra extended by $\kappa$-Minkowski noncommutative
coordinates. Therefore, introduced realizations for Lorentz algebras are compatible
with deformed Heisenberg algebras as well.

One can consider the following ansatz for generators in which the deformation
parameter $a$ has the form $(a_0,0,0,0)$ and the rotation subalgebra \eso(3)
is undeformed: 
\begin{equation}\label{ansatzM}
\hM_{i0}=x_{i}p_{0}F_{1}\left(A,b\right) -x_{0}p_{i}F_{2}\left( A,b\right)
+a_{0}\left( x_{k}p_{k}\right) p_{i}F_{3}\left( A,b\right) +a_{0}x_{i}\vec{p}%
^{2}F_{4}\left( A,b\right) 
\end{equation}
\begin{eqnarray}
\hM_{ij}=M_{ij}=x_{i}p_{j}-x_{j}p_{i},
  \label{c10}
\end{eqnarray}
expressed in terms of undeformed Heisenberg algebra, and $A=a_0p_0;b=a_{0}^{2}
\vec{p}^{2}$. When $a_0$ goes to 0, $F_1$ and $F_2$ go to 1, $F_3$ and $F_4$
are finite and $g_{00}$ goes to $\eta_{00}$.

The generalized Poincar\'{e} algebra is described by the
following set of commutation relations:
\begin{equation}
\left[ \hM_{i0},p_{0}\right] =ip_{i}F_{2}; \qquad \left[
\hM_{i0},p_{j}\right] =i\delta _{ij}\left( p_{0}F_{1}+a_{0}\vec{p}%
^{2}F_{4}\right) +ia_{0}p_{i}p_{j}F_{3}
\label{c13}
\end{equation}
\begin{equation}\label{c14}
  [M_{ij},\hM_{k0}]=i\left(\hM_{j0}\eta_{ik}-\hM_{i0}\eta_{jk}\right)
\end{equation}
\begin{eqnarray}\label{gen_alg}
\left[ \hM_{i0},\hM_{j0}\right]
=i\hM_{ij}(-F_{1}F_{2}+AF_{1}F_{3}-2AF_{1}F_{4}-bF_{3}F_{4}-
2bF_{4}^{2}-A\frac{%
\partial F_{1}}{\partial A}F_{2}-bF_{2}\frac{\partial F_{4}}{\partial A}\\
\nonumber
-2A^{2}F_{1}\frac{\partial F_{1}}{\partial b}-2Ab\frac{\partial F_{1}}{%
\partial b}F_{3}-2Ab\frac{\partial F_{1}}{\partial b}F_{4}-2AbF_{1}\frac{%
\partial F_{4}}{\partial b}-2b^{2}F_{3}\frac{\partial F_{4}}{\partial b}%
-2b^{2}F_{4}\frac{\partial F_{4}}{\partial b}),
\end{eqnarray}
(the rest of commutation relations for Poincar\'{e} algebra stays undeformed). Note
that Eqs. (\ref{ansatzM})-(\ref{gen_alg}) also include the case of the
undeformed Lorentz algebra.

In the classical limit $a_0\shortrightarrow 0$ from 
generalized Poincar\'{e} algebra
we recover the well-known relations of the undeformed (standard) Poincar\'{e} algebra
$\mathcal{U}_{\mathrm{iso}(1,3)}$ : $g_{\mu\nu}\rightarrow\eta_{\mu\nu}$ and
$G_{\mu\nu\lambda}(p)\rightarrow i\left(p_\nu\eta_{\mu\lambda}-p_\mu
\eta_{\nu\lambda}\right)$.
One can easily recover the form of the metric $g_{\mu\nu}(p)$ for which the set of
commutation relations (\ref{c13})-(\ref{gen_alg}) is satisfied. It leads to the following:
\begin{eqnarray}
g_{00}(p)=-F_{1}F_{2}+AF_{1}F_{3}-2AF_{1}F_{4}-bF_{3}F_{4}-
2bF_{4}^{2}-A\frac{%
\partial F_{1}}{\partial A}F_{2}-bF_{2}\frac{\partial F_{4}}{\partial A}
-2A^{2}F_{1}\frac{\partial F_{1}}{\partial b}\\
-2Ab\frac{\partial F_{1}}{%
\partial b}F_{3}-2Ab\frac{\partial F_{1}}{\partial b}F_{4}-2AbF_{1}\frac{%
\partial F_{4}}{\partial b}-2b^{2}F_{3}\frac{\partial F_{4}}{\partial b}%
-2b^{2}F_{4}\frac{\partial F_{4}}{\partial b} \nonumber
\end{eqnarray}
and $g_{ij}=\eta_{ij}$,  $g_{i0}=0$.\footnote{
Within the idea of relative locality \cite{rel_loc} the metric dependence on momenta
is linear, here we have more general form, however also consistent with the
$\kappa$-Poincar\'{e} algebra which will be shown in Section \ref{dla}.}

Remark: It is possible to define Hermitian realization of Lorentz
generators as well \footnote{Note that if \hMmn and $p_\rho$
are Hermitian then the right hand side of (\ref{c4}) should be anti-Herimitian.
Also $g_{00}(A,b)$ commutes with $M_{ij}$ in (\ref{c14}).}:
\begin{equation}
  \hat{M}_{i0}^{h}=\frac{\hat{M}_{i0}+\hat{M}_{i0}^{\dagger }}{2}.
\end{equation}

The deformed symmetry algebra can be equipped in Hopf algebra
structure as well, generally $\igel (4)$ one.
As quantum Hopf algebra the generalized Poincar\'{e} algebra will possess
an algebraic sector (commutators in the form of (\ref{c13})-(\ref{gen_alg}))
and the co-algebraic sector (coproduct, counit and antipode), generally in $\igel (4)$.
The quantum (deformed) Hopf algebras will be presented in the next sections.
Let us mention that when $a_0$  goes to 0 it reduces to undeformed Hopf algebra,
(\ref{isog1})-(\ref{SPoin}).

\section{Lorentz algebra and \ka-Poincar\'{e}-Hopf algebra}
\label{lapha}

In this section we consider the Lie algebra defined by undeformed Lorentz
generators \Mmn\ and $\hat{x}_\mu$ (\ref{b3}),
(see \cite{k-gn02,glik02,k-jm11,km11,ms06,k-jms07,bp11,bp10}) satisfying
the following commutation relation
\begin{equation}
  [M_{\mu\nu},\hat{x}_\lambda]=i\left(
    \hat{x}_\nu\eta_{\mu\lambda}-\hat{x}_\mu\eta_{\nu\lambda}+
    a_\nu M_{\mu\lambda}-a_\mu M_{\nu\lambda}\right).
  \label{d2}
\end{equation}
In this Lie algebra we use $M_{\mu\nu}$ instead of $\hat{M}_{\mu\nu}$.
Let us mention that the relation (\ref{d2}) is the unique way to obtain a Lie
algebra generated by \Mmn\ and $\hat{x}_\mu$ (see \cite{k-jms07}).
For a class of Heisenberg algebras defined in Section \ref{dha} by
$h_{\mu\nu}(p)$, we get the corresponding class of \ka-deformed Poincar\'{e}
algebras defined by
\begin{equation}\label{d6}
  [\Mmn,p_\lambda]=iG_{\mu\nu\lambda}(p).
\end{equation}
Jacobi identities imply that $G_{\mu\nu\lambda}$ is uniquely determined by
$h_{\mu\nu}$ \cite{km11}.

One example is particularly interesting: the {\itshape natural} realization (or
{\itshape classical bases}) \cite{k-gn02,k-jms07,bp10,k-jm11,km11}.
For the {\itshape natural} realization,
\begin{eqnarray}
  &&\hat{x}_\mu=X_\mu Z^{-1}+a_0X_0P_\mu\nonumber\\
  &&\Mmn=X_\mu P_\nu-X_\nu P_\mu, \ \ (F_{1}=F_{2}=1, F_{3}=F_{4}=0),
  \label{d1}
\end{eqnarray}
where $Z^{-1}=-a_0P_0+\sqrt{1-a_0^2P^2}$. In order to emphasize the
{\itshape natural} realization, we write capital letters (to distinguish from generic
$x_{\mu}$ and $p_{\mu}$). Then the commutator of \Mmn\ and $P_\lambda$ has
the form
\begin{equation}
  [\Mmn,P_\lambda]=i\left(P_\nu\eta_{\mu\lambda}-P_\mu\eta_{\nu\lambda}\right).
\end{equation}
They generate the undeformed Poincar\'{e} algebra.
The coalgebra structure is given by
\begin{eqnarray}
  &&\Delta P_\mu=P_\mu\otimes Z^{-1}+1\otimes P_\mu-a_\mu p_\alpha^L
      Z\otimes P^\alpha\nonumber\\
  &&\Delta\Mmn=\Mmn\otimes 1+1\otimes\Mmn-a_\mu (p^L)^\alpha Z
      \otimes M_{\alpha\nu}+a_\nu (p^L)^\alpha Z\otimes M_{\alpha\mu}
\end{eqnarray}
where
\begin{equation}
  p_\alpha^L=P_\alpha-\frac{a_\alpha}2\Box
\end{equation}
and $\Box=\frac2{a^2}
\left(1-\sqrt{1+a^2P^2}\right)$ ($a^2=a_\alpha a^\alpha$,
$P^2=P_\alpha P^\alpha$).
Here, the deformation is described by the general four-vector $a_\mu$.
Counits are trivial i.e. $\epsilon(\Mmn)=0$ and $\epsilon(P_\mu)=0$. The antipodes
are given by
\begin{eqnarray}
  &&S(P_\mu)=\left(-P_\mu-a_\mu p^LP\right)Z\nonumber\\
  &&S(\Mmn)=-\Mmn-a_\mu p_\alpha^L M_{\alpha\nu}+a_\nu
    p_\alpha^LM_{\alpha\mu}.
\end{eqnarray}

It is important to emphasize that using similarity transformations
\begin{eqnarray}
  P_\mu&=&\mE p_\mu\mE^{-1}=P_\mu(p)\label{d3}\\
  X_\mu&=&\mE x_\mu\mE^{-1}=x^\alpha\Psi_{\alpha\mu}(p)\label{d4}
\end{eqnarray}
where $\mE=\exp\{x^\alpha\Sigma_\alpha(p)\}$ and functions
$\ds \Sigma_\alpha(p)$ satisfy the boundary condition
\begin{equation}
  \lim_{a\rightarrow0}\Sigma_\alpha=0,
\end{equation}
one can obtain formulae for the coproduct and antipode of $p_\mu$ using
(\ref{d3}) (and its inverse relation) \cite{km11}.

Let us consider the class of realizations given by (\ref{real}).
The corresponding similarity transformations for $P_\mu$ (\ref{d3}) are
given by
\begin{eqnarray}
  &&P_0=\frac{1-Z^{-1}}{a_0}+\frac{a_0}2\Box\\
  &&P_i=p_i\frac{Z^{-1}}{\varphi(A)},
\end{eqnarray}
where
\begin{equation}
  \Box=-\vec{p}^2\frac{Z^{-1}}{\varphi^2(A)}+\frac1{a_0^2}(Z+Z^{-1}-2).
\end{equation}
It is easy to obtain formulae for $X_0$ and $X_i$ \cite{k-jm11}.

Functions $F_i(A,b)$ are given by
\begin{equation}
  F_1(A,b)=\frac{\varphi(A)(Z^2-1)}{2A},\;\;
  F_2(A,b)=\frac{\psi(A)}{\varphi(A)},\;\;
  F_3(A,b)=\frac{\gamma(A)}{\varphi(A)},\;\;
  F_4(A,b)=-\frac{1}{2\varphi(A)}. \label{Fi8}
\end{equation}

Let $\ds \Psi(A)=\int_0^A\frac{{\mathrm d}t}{\psi(t)}$.
The coalgebra structure is given by
\begin{equation}
  \Delta p_0=\frac1{a_0}\Psi^{-1}\left(\ln(Z\otimes Z)\right)
\end{equation}
where $\ln(Z\otimes Z)=\ln(Z)\otimes1+1\otimes\ln(Z)$. Also
\begin{equation}
  \Delta p_i=\varphi(a_0\Delta p_0)\left(\frac{p_i}{\varphi(a_0p_0)}\otimes1+
    Z\otimes\frac{p_i}{\varphi(a_0p_0)}\right).
\end{equation}
The coproducts of the Lorentz generators are given by
\begin{equation}
  \Delta M_{i0}=M_{i0}\otimes1+Z\otimes M_{i0}-a_0
      \frac{p_j}{\varphi(a_0p_0)}\otimes M_{ij}
  \label{d7}
\end{equation}
and
\begin{equation}
  \Delta M_{ij}=M_{ij}\otimes1+1\otimes M_{ij}.
\end{equation}
The counits of all generators are undeformed and the antipodes are given by
\begin{eqnarray}
  &&S(p_0)=\frac1{a_0}\Psi^{-1}(\ln(Z^{-1})),\\
  &&S(p_i)=-p_i\frac{\varphi(S(a_0p_0))}{\varphi(a_0p_0)}Z^{-1},\\
  &&S(M_{i0})=-Z^{-1}\left(M_{i0}+a_0\frac{p_j}{\varphi(a_0p_0)}M_{ij}\right),\\
  &&S(M_{ij})=-M_{ij}.
\end{eqnarray}

For the choice $\psi=1$ and $\varphi=Z^{-\lambda}$ in (\ref{real})
the Hopf algebra structure can be written as:
\begin{eqnarray}\label{copP}
\Delta p_{0} =p_0\otimes1+1\otimes p_0; \qquad\Delta
p_{i} =p_{i}\otimes Z^{-\lambda }+Z^{1-\lambda }\otimes p_{i};\\
\Delta M_{i0}=M_{i0}\otimes1+Z\otimes M_{i0}-a_0Z^\lambda
      p_j\otimes M_{ij}\label{d8}
\end{eqnarray}
\begin{equation}
  S(p_0)=-p_0;\qquad S(p_i)=-Z^{2\lambda-1}p_i;\qquad
  S(M_{i0})=-Z^{-1}\left(M_{i0}+a_0Z^\lambda p_jM_{ij}\right).
  \label{d9}
\end{equation}

Let us consider two choices of $\lambda$:\\
1) The case $\lambda =0$.\\
The form of functions $F_i$ is the following:
$\ds F_1=\frac{Z^2-1}{2A},\;F_2=1,\;F_3=1,\;F_4=-\frac{1}{2}$.
The formulae for the coproduct  and antipode of momentum and Lorentz
generators are written in formulae (\ref{copP}), (\ref{d8}) and (\ref{d9}).
The Hopf algebra obtained in this way is identical to the
bicrossproduct basis from \cite{MR}.\\
2) The case $\lambda=\frac{1}{2}$.\\
The form of functions $F_i$ is the following:
$F_1=\frac{\sinh A}{A}Z^\frac12,\;F_2=Z^\frac12,\;F_3=\frac12Z^\frac12,\;
F_4=-\frac12Z^\frac12$.
Again, the formulae for the coproduct  and antipode of momentum and Lorentz
generators are written in formulae (\ref{copP}), (\ref{d8}) and (\ref{d9}).
Coproducts and antipodes for momenta, $p_0$, $p_i$ are identical as in the
standard basis \cite{Luk1}. Since the Lorentz algebra is undeformed, the
coproducts of $M_{i0}$ are not the same as in the standard basis \cite{Luk1}.

Similar examples could be presented for
$\psi=1$ and $\varphi=\frac{A}{e^A-1}$ ({\itshape Weyl symmetric} realization),
$\psi=\varphi=1-A$ ({\itshape left covariant} realization) and
$\psi=1+A$ and $\varphi=1$ ({\itshape right covariant} realization).
There are some other interesting examples in \cite{km11}.

The generalized form of the Poincar\'{e} algebra does not contain dilatation
generator, however the ansatz for Lorentz boosts includes them (see formula
(\ref{ansatzM})). Nevertheless provided examples show that Poincar\'{e} algebra
does not necessarily need to be extended, even in the case of $F_3\neq 0$.

We point out that for the given choice of $h_{\mu\nu}$, there is
a unique expression for $G_{\mu\nu\lambda}$ (\ref{d6}) and $F_i$ (\ref{ansatzM})
such that the Lorentz algebra is undeformed and relation (\ref{d2}) is satisfied.
Note that the reverse statement is not true.
Examples with undeformed Poincar\'{e} algebras are given by a
family of $h_{\mu\nu}$, see Subsection 4.2. in \cite{km11}.

\section{\ka-Deformed Lorentz algebras and \ka-Poincar\'{e}-Hopf algebra}
\label{dla}

Let us consider the given choice of $h_{\mu\nu}$ (Eq. (\ref{b3})) and a
family of \ka-deformed Lorentz algebras defined by functions $F_1^d$,
$F_2^d$, $F_3^d$ and $F_4^d$ (see (\ref{ansatzM}) and (\ref{c10})).
For a given choice of $h_{\mu\nu}$, there is a unique choice of functions $F_i$
(denoted without index $d$), such that the Lorentz algebra is undeformed and
Lorentz generators \Mmn\ and noncommutative coordinates $\hat{x}_\mu$
form a Lie algebra via the relation (\ref{d2}) (Section \ref{lapha}). Now, let us
restrict our attention to the family of \ka-deformed Lorentz algebras which are
related to the undeformed one by the relation
\begin{equation}
  \hM_{i0}=M_{j0}(\delta_{ij}G_1+a_0^2p_ip_jG_2)+a_0M_{ij}p_jG_3
  \label{e4}
\end{equation}
where $G_i(A,b)$ are arbitrary functions such that $G_1$ goes to 1 and
$G_2$ and $G_3$ are finite as the deformation parameter $a_0$ goes
to 0. We emphasize that (\ref{e4}) is a subclass of (\ref{ansatzM}).
For simplicity we consider the case $G_2= 0$ and (\ref{e4}) transforms to
\begin{equation}
  \hM_{i0}=M_{i0}G_1+a_0M_{ij}p_jG_3.
  \label{e7}
\end{equation}
The inverse relation is given by
\begin{equation}
  M_{i0}=\hM_{i0}\frac1{G_1}-a_0M_{ij}p_j\frac{G_3}{G_1}.
  \label{e8}
\end{equation}
Relations connecting $F_i^d$ and $F_i$ are given by
\begin{equation}
  F_1^d=F_1G_1,\;\;\;F_2^d=F_2G_1,\;\;\;F_3^d=F_3G_1-G_3,\;\;\;
  F_4^d=F_4G_1+G_3.
\end{equation}
It is easy to obtain inverse relations.
Let us mention that generally $g_{00}\neq\eta_{00}$ ((\ref{c4}) and
(\ref{gen_alg})), implying that the Lorentz algebra is deformed. If $G_1=1$ and
$G_3=0$, then $g_{00}=\eta_{00}$.

The coproduct $\Delta \hM_{i0}$ can be obtained from Eqs. (\ref{e7})
and (\ref{e8}):
\begin{equation}
  \Delta\hM_{i0}=\Delta(M_{i0})\Delta(G_1)+a_0\Delta(M_{ij})\Delta(p_j)
    \Delta(G_3).
  \label{e10}
\end{equation}
Coproducts $\Delta p_i$ and $\Delta M_{i0}$ are calculated in the previous
section. One has to express $M_{i0}$ in terms of
$\hM_{i0}$ and $M_{ij}$ (\ref{e8}) in the Eq. (\ref{e10}).

For the class of examples when $\psi=1$ and $\varphi=Z^{-\lambda}$
the relation (\ref{e10}) transforms to
\begin{eqnarray}
  \Delta\hM_{i0}&=&\left(M_{i0}\otimes1+Z\otimes M_{i0}-a_0
    p_j Z^\lambda\otimes M_{ij}\right)\Delta G_1\nonumber\\
    &+&a_0\left(M_{ij}\otimes1+1\otimes M_{ij}\right)
    \left(p_{j}\otimes Z^{-\lambda }+Z^{1-\lambda }\otimes p_{j}\right)
    \Delta G_3\nonumber\\
    &=&\left(\hM_{i0}\frac1{G_1}\otimes1+Z\otimes\hM_{i0}\frac1{G_1}
    -a_0\left(M_{ij}p_j\frac{G_3}{G_1}\otimes1+Z\otimes M_{ij}p_j\frac{G_3}
    {G_1}+p_j Z^\lambda\otimes M_{ij}\right)\right)\Delta G_1
    \nonumber\\
    &+&a_0\left(M_{ij}\otimes1+1\otimes M_{ij}\right)
    \left(p_{j}\otimes Z^{-\lambda }+Z^{1-\lambda }\otimes p_{j}\right)
    \Delta G_3.
\end{eqnarray}

Let us consider the special case $\lambda=\frac12$,
$\ds F_1^d=\frac{\sinh A}A$, $F_2^d=1$, $F_3^d=F_4^d=0$, $G_1=
Z^{-\frac12}$ and $G_3=\frac12$:
\begin{equation}
  \hM_{i0}=M_{i0}Z^{-\frac12}+\frac12a_0M_{ij}p_j
\end{equation}
and
\begin{equation}
  \Delta\hM_{i0}=\hM_{i0}\otimes Z^{-\frac12}+Z^\frac12\otimes\hM_{i0}+
    \frac{a_0}2\left(M_{ij}Z^\frac12\otimes p_j-p_j\otimes M_{ij}Z^{-\frac12}\right).
\end{equation}
Let us mention that $g_{00}=\eta_{00}\cosh A$.
This example for $\hM_{i0}$ and $\Delta\hM_{i0}$ coincides with the corresponding
results in the standard basis in \cite{Luk1} ($a_0=-\frac1\kappa$).

Eq. (\ref{e4}) produces a three-parameter ($G_1,\, G_2,\, G_3$) family of generalized Poincar\'{e} algebras,
in which the Lorentz algebra is generally deformed. This family corresponds to different bases of the $\kappa$-Poincar\'{e}-Hopf algebra. We point out that for all the other deformed Lorentz generators $\hM_{i0}$, which
are not related to \Mmn\ via relation (\ref{e4}), $\hM_{i0}$ can be related to some undeformed Poincar\'{e} algebra, but not to the Poincar\'{e} algebra given by the {\itshape natural} realization (\ref{d1}). As a consequence, the coproduct of such $\hM_{i0}$ cannot be written in terms of (deformed) Poincar\'{e} generators only, the Hopf algebra structure should be extended to $\kappa$-deformed $\mathfrak{igl}(4)$ (or Poincar\'{e}-Weyl) Hopf algebra. Note that undeformed Lorentz generators $\widetilde{\Mmn}=x_\mu p_\nu-x_\nu p_\mu$ generally cannot be written in terms of \Mmn\ by the Eq. (\ref{e4}). For example
\begin{equation}
M_{i0} =X_i P_0-X_0 P_i 
=\Biggl(x_{i}p_{0}\frac{sinh A}{A} -x_{0}p_{i} +\frac{a_{0}}{2}\left( x_{k}p_{k}\right) p_{i} -\frac{a_{0}}{2}x_{i}\vec{p}%
^{2}\Biggr)Z^{\frac{1}{2}},
\end{equation}
for the case $\lambda=\frac{1}{2}$ in Section \ref{lapha}. Inserting generators $\widetilde{M}$ and $M$ into relation (\ref{e4}), we find that there is no solution for $G_1,\, G_2,\, G_3$. As a consequence, $\Delta \widetilde{M}_{i0}$ is not closed in $\U_{\mathfrak{iso}(1,3)}$, which we demonstrate in Section \ref{mst}.

\section{\ka-Minkowski spacetime and twisting $\igel (4)$}
\label{mst}

To each realization (\ref{real}), there is a corresponding twist. In \cite{gghmm08},
Abelian twists for realizations (\ref{real}) where $\psi=1$ were constructed.
In \cite{bp09}, Jordanian twists for one subfamily of realizations (\ref{real})
with $\psi\neq1$ were constructed.

Let us consider the (undeformed) algebra \igel(4) generated
by $L_{\mu\nu}= x_\mu p_\nu$ and $p_\mu$.
The commutation relations are given by
\begin{equation}\label{f1}
  [L_{\mu\nu},L_{\lambda\rho}]=-i\left(L_{\mu\rho}\eta_{\nu\lambda}-
    L_{\lambda\nu}\eta_{\mu\rho}\right);\;\;\;
  [L_{\mu\nu},p_\lambda]=i\eta_{\mu\lambda}p_\nu.
\end{equation}
The commutation relations $[L_{\mu\nu},\hat{x}_\lambda]$ can be easily
calculated using (\ref{real}) and (\ref{f1}).

The coalgebra structure can be obtained by the twist operator $\mathcal{F}$.
For the given twist operator,
$\Delta p_\mu=\mathcal{F}\Delta_0p_\mu\mathcal{F}^{-1}$ and
$\Delta L_{\mu\nu}=\mathcal{F}\Delta_0L_{\mu\nu}\mathcal{F}^{-1}$
($\Delta_0p_\mu$ and $\Delta_0L_{\mu\nu}$ are primitive coproducts),
while the counit is trivial.

For the realizations (\ref{real}) given by $\psi=1$ and $\varphi=Z^{-\lambda}$,
the corresponding Abelian twists \cite{gghmm08,bp09} are
\begin{equation}
  \mathcal{F}=\exp\left(\lambda ix_kp_k\otimes A-(1-\lambda)A
    \otimes ix_kp_k\right).
  \label{f3}
\end{equation}
Coproducts of $p_\mu$, obtained by twist (\ref{f3}) coincide with
Eq. (\ref{copP}).

Generators $\widetilde{M_{\mu\nu}}=L_{\mu\nu}-L_{\nu\mu}$ generate the
subalgebra \eso(1,3) of \gel(4).
It can be shown that 
\begin{eqnarray}
  \Delta\widetilde{M_{i0}}&=&
    L_{i0}\otimes Z^{\lambda}+Z^{-(1-\lambda)}\otimes L_{i0}\\
    &-&L_{0i}\otimes Z^{-\lambda}-Z^{1-\lambda}\otimes L_{0i}-
    (1-\lambda)a_0p_i\otimes\sum_{k=1}^3L_{kk}Z^{-\lambda}+
    \lambda\sum_{k=1}^3L_{kk}Z^{1-\lambda}\otimes a_0p_i.\nonumber
\end{eqnarray}
Hence, $\Delta\widetilde{M_{i0}}$ cannot be expressed in terms of Poincar\'{e}
generators $\widetilde{M_{\mu\nu}}$ and $p_\mu$. It shows that it is not possible
to put the coalgebra structure on the subalgebra \ieso(1,3), and we get
\ka-deformed \igel(4)\ Hopf algebra compatible with \ka-Minkowski spacetime.

Similarly, one can obtain $\Delta L_{\mu\nu}$ and $\Delta p_\mu$ using the
family of Jordanian twists \cite{bp09}. Particularly, for {\itshape left covariant}
($\psi=\varphi=1-A$), and {\itshape right covariant} ($\psi=1+A,\varphi=1$)
realizations, corresponding Jordanian twists lead to the Poincar\'{e}-Weyl algebra
which includes dilatations \cite{byk09}.

We point out that for the given $h_{\mu\nu}$, there are infinitely many ways of
implementing \ka-deformed \gel(4) or \ka-deformed Lorentz algebras defined by
Eqs. (\ref{ansatzM}) and (\ref{c10}).

\section{Conclusions}
\label{conc}

The structure of spacetime, at the scale where quantum gravity effects take place,
is one of the most important questions in fundamental physics. One of the examples
of noncommutative structure is $\kappa$-Minkowski spacetime. Below the quantum
gravity scale the symmetry of spacetime should also be deformed. In this Letter we
have generalized the description of such deformations, which include the well-known
forms of $\kappa$-Poincar\'{e}-Hopf algebra in different basis as special cases.

For the given phase space $\hat{H}$, defined by $h_{\mu\nu}$ (\ref{b3}), there
exists a unique undeformed Lorentz algebra such that $\Mmn$ and
$\hat{x}_\lambda$ generate the Lie algebra (\ref{d2}). Then, the realization of
Lorentz generators is fixed. For example, they are given by (\ref{ansatzM}),
(\ref{c10}) and (\ref{Fi8}), corresponding to the family of realizations (\ref{real}).
The relations between $M_{i0}$ and $p_{\mu}$ are given by (\ref{c13}) and
(\ref{Fi8}). The Hopf algebra structure is fixed and it corresponds to the
$\ka$-Poincar\'{e}-Hopf algebra with undeformed Lorentz algebra \cite{km11}.

For the given $h_{\mu\nu}$ we have constructed a three-parameter family of
deformed Lorentz generators $\hat{M}_{i0}$, described by $G_{1}$, $G_2$ and
$G_{3}$ (\ref{e4}). The generators $\hat{M}_{i0}$ are given by (\ref{ansatzM})
and (\ref{c10}). The Hopf algebra structure $\Delta \hat{M}_{i0}$ depends on
$G_{1}$ and $G_{3}$ (\ref{e10}). The $\kappa$-Poincar\'{e}-Hopf algebra with
deformed Lorentz algebra is related to the $\kappa$-Poincar\'{e}-Hopf algebra with
undeformed Lorentz algebra. A special example where $\lambda=\frac12$, $G_1=
Z^{-\frac12}$ and $G_3=\frac12$ coincides with the standard basis given in
\cite{Luk1}.

For the same choice of $h_{\mu\nu}$ there is an infinite family of deformed Lorentz
generators $\hat{M}_{i0}$ given by $F_i$ (\ref{ansatzM}) such that the
corresponding Hopf algebra structure is not \ka-Poincar\'{e}, but lies in $\igel (4)$.
For example, $\tilde{M}_{\mu\nu}=x_\mu p_\nu-x_\nu p_\mu,\;
(F_1=F_2=1,\;F_3=F_4=0)$ does have the coproduct in $\igel (4)\otimes
\igel(4)$ (for generic $h_{\mu\nu}$ see Section \ref{mst}). Note that for the
{\itshape natural} realization (\ref{d1}) generators $M_{\mu\nu}$ have the
\ka-Poincar\'{e}-Hopf algebra structure (see Section \ref{lapha}). 

We point out that for the given phase space $\hat{H}$ defined by $h_{\mu\nu}$,
there are infinitely many implementations of \ka-deformed Poincar\'{e} (and
$\igel (4)$) algebras and corresponding Hopf algebras. Our description
contains them both (\ka-deformed Poincar\'{e} and
$\igel (4)$) therefore we call it generalized. There is no physical principle
which would distinguish some choice in this class. What are good and bad physical
consequences of some choice (which generally leads to Lorentz symmetry violation)
is still an open question.

Nevertheless such a general form for the $\kappa$-Poincar\'{e}-Hopf algebra might
be useful in the view for applications in the Planck scale or Quantum Gravity physics.
Certain realizations of quantum spacetime influence the form of the symmetry algebra
and deform the Casimir operators, which in turn lead to different dispersion relations.
As it was already shown (see, e.g., \cite{bgmp10}) such dispersion relations result in
the so-called time delays for photons, which could be connected with the similar effect
measured for high energy photons coming from gamma ray bursts (GRB's). 

\section*{Acknowledgements}

The authors would like to gratefully acknowledge helpful comments from A. Borowiec,
J. Lukierski and A. Samsarov. This work was supported by the Ministry of Science and
Technology of the Republic of Croatia under contract No. 098-0000000-2865.
A.P. acknowledges the financial support of Polish NCN grant No. 2011/01/B/ST2/03354.


\begin{thebibliography}{99}

\bibitem{Luk1}
  J. Lukierski, A. Nowicki, H. Ruegg, V. N. Tolstoy,
  \textit{
  Phys. Lett.} \textbf{B264}, 331 (1991);

  J. Lukierski, H. Ruegg,
  \textit{Phys. Lett.} \textbf{B329}, 189 (1994),
  [arXiv:hep-th/9310117].

\bibitem{MR}
  S. Majid, H. Ruegg,
  \textit{Phys. Lett.} \textbf{B334}, 348 (1994),
  [arXiv:hep-th/9405107].
  
\bibitem{Dop1}
  S. Doplicher, K. Fredenhagen, J. E. Roberts,
  \textit{Phys. Lett.} \textbf{B331}, 39 (1994);
  S. Doplicher, K. Fredenhagen, J. E. Roberts,
  \textit{Commun. Math. Phys.} \textbf{172}, 187 (1995),
  [arXiv:hep-th/0303037].

\bibitem{klm00}
  P. Kosinski, J. Lukierski, P. Maslanka,
  \textit{Phys. Rev.} \textbf{D 62} 025004 (2000),
  [arXiv:hep-th/9902037];
  M. Daszkiewicz,  J. Lukierski, M. Woronowicz,
  \textit{Phys. Rev.} \textbf{D77}, 105007  (2008),
  [arXiv:hep-th/0708.1561].

\bibitem{klry09}
    {
      H. C. Kim, Y. Lee, C. Rim and J. H. Yee},
    J. Math. Phys. {\bfseries 50} 102304 (2009),
    [arXiv:hep-th/0901.0049].

\bibitem{ms11}
  S. Meljanac, A. Samsarov,
  \textit{Int. J. Mod. Phys.} \textbf{A26} 1439-1468 (2011),
  [arXiv:1007.3943].

\bibitem{mstw11}
  S. Meljanac, A. Samsarov, J. Trampetic, M. Wohlgenannt,
  \textit{JHEP} \textbf{1112} 010 (2011),
  [arXiv:1111.5553].

\bibitem{hjm11}
  E. Harikumar, T. Juric, S. Meljanac,
  \textit{Phys. Rev.} \textbf{D84} 085020 (2011),
  [arXiv:1107.3936];
  E. Harikumar, \textit{Europhys. Lett.} \textbf{90} 21001 (2010),
  [arXiv:1002.3202v3].

\bibitem{dj11}
  M. Dimitrijevic, L. Jonke,
  \textit{JHEP} \textbf{1112} 080 (2011),
  [arXiv:1107.3475].

\bibitem{kappaSt}
  M. Daszkiewicz, J. Lukierski and M. Woronowicz,
  Mod. Phys. Lett. A{\bfseries 23} 653-665 (2008),
  [arXiv:hep-th/0703200];
  M. Arzano and A. Marciano,
  Phys. Rev. \textbf{D76}, 125005 (2007),
  [arXiv:hep-th/0701268];
  C. A. S. Young and R. Zegers,
  Nucl. Phys. \textbf{B797}, 537 (2008),
  [arXiv:0711.2206];
  C. A. S. Young and R. Zegers,
  Nucl. Phys. \textbf{B804}, 342 (2008),
  [arXiv:0803.2659].

\bibitem{gghmm08}
    {
      T. R. Govindarajan, K. S. Gupta, E. Harikumar, S. Meljanac and
        D. Meljanac}, Phys. Rev. {\bfseries  D 77} 105010 (2008),
    [arXiv:hep-th/0802.1576];
   T. R. Govindarajan, Kumar S. Gupta, E. Harikumar, S. Meljanac, D. Meljanac,
   \textit{Phys. Rev.} \textbf{D80} 025014 (2009), [arXiv:0903.2355v3].

\bibitem{k-jm11}
  S. Meljanac, S. Kresic-Juric,
  Int. J. Mod. Phys. \textbf{A 26} (20) (2011), 3385-3402,
  [arXiv:1004.4647].

\bibitem{km11}
  {
    D. Kova\v{c}evi\'{c} and S. Meljanac}, \textit{Kappa-Minkowski spacetime, Kappa-Poincare-Hopf algebra and realizations}
  [arXiv:math-ph/1110.0944].
  
\bibitem{glik02}
  J. Kowalski-Glikman,
  \textit{Phys. Lett.} \textbf{B 547}, 291 (2002),
  [arXiv:hep-th/0207279].

\bibitem{k-gn02}
   J. Kowalski-Glikman, S.Nowak,
  \textit{Phys. Lett.} {\bf B 539}, 126 (2002)
  [arXiv:hep-th/0203040];
  L. Freidel, J. Kowalski-Glikman, S. Nowak,
  \textit{Int. J. Mod. Phys.} \textbf{A 23}, 2687 (2008),
  [arXiv:0706.3658].

\bibitem{bklvy}
    {
      J. G. Bu, H. C. Kim, Y. Lee, C. H. Vac and J. H. Yee},
    Phys. Lett.{\bfseries  B  665} 95-99 (2008),
    [arXiv:hep-th/0611175].

\bibitem{klry08}
    {
      H. C. Kim, Y. Lee, C. Rim and J. H. Yee},
    Phys. Lett.  {\bfseries  B  671} 398-401 (2009),
    [arXiv:hep-th/0808.2866].

\bibitem{bp09}
    {
      A. Borowiec and A. Pacho\l},
    Phys. Rev. {\bfseries D 79} 045012 (2009),
    [arXiv:math-ph/0812.0576].

\bibitem{byk09}
    {
      J. G. Bu, J. H. Yee and H. C. Kim},
    Phys. Lett.  {\bfseries B 679} 486-490 (2009),
    [arXiv:hep-th/0903.0040].

\bibitem{k-jms07}
    S. Meljanac, S. Kresic-Juric and M. Stojic,
    \textit{Eur.Phys.J.} \textbf {C 51}, 229-240 (2007),
    [arXiv:hep-th/0702215].

\bibitem{ms06}
    S. Meljanac, M. Stojic,
    \textit{Eur. Phys. J.} \textbf{C 47} 531-539 (2006),
    [arXiv:hep-th/0605133];
    S. Meljanac, A. Samsarov, M. Stoji\'{c}, K. S. Gupta,
    \textit{Eur. Phys. J.} \textbf{C 53}, 295 (2008),
    [arXiv:0705.2471].

\bibitem{bp11}
    A. Borowiec, A. Pacho\l,
    \textit{SIGMA} \textbf{6}, 086 (2010),
    [arXiv:1005.4429].

\bibitem{bp10} A. Borowiec, A. Pacho\l,
    \textit{J. Phys.} \textbf{A 43}, 045203 (2010),
    [arXiv:0903.5251].

\bibitem{bgmp10}
    A. Borowiec, Kumar S. Gupta, S. Meljanac A. Pacho{\l},
    \textit{Europhys. Lett.} \textbf{92}, 20006 (2010),
    [arXiv:0912.3299].
\bibitem{rel_loc}
G. Amelino-Camelia, M. Arzano, J. Kowalski-Glikman, G. Rosati, G. Trevisan
\textit{Relative-locality distant observers and the phenomenology of momentum-space geometry}
[arXiv:1107.1724].
%
\end{thebibliography}
\end{document}